\documentclass[11pt]{article}
\usepackage{times}
\usepackage{sidecap_mod,xspace}
\usepackage{geometry}
\usepackage{amsmath}
\usepackage{natbib,bm}
\usepackage{floatrow}
\usepackage{float}
\usepackage{hyperref}
\usepackage{graphics,graphicx}
\usepackage{subcaption}
\usepackage{epsf,wrapfig}
\usepackage[symbol]{footmisc}

\usepackage{paralist}
\let\olditem\item
\renewenvironment{thebibliography}[1]{%
  \section*{\refname}
  \let\par\relax
  \renewcommand{\item}[1][]{\olditem[\textbullet]}%
  \inparaenum}{\endinparaenum}

\newcounter{aff}

\DeclareGraphicsExtensions{.jpg,.pdf,.png,.eps}

\usepackage{caption}
\usepackage{xcolor}

\usepackage{tcolorbox}

\bibpunct{(}{)}{;}{a}{}{,}

\usepackage{soul}

\definecolor{darkcyan}{rgb}{0.0, 0.0, 0.7}

\definecolor{boxcolor}{RGB}{193, 229, 255}
\definecolor{figboxcolor}{RGB}{193, 229, 255}
\definecolor{boxcolorgray}{RGB}{227, 227, 227}
\definecolor{darkcyan}{rgb}{0.0, 0.0, 0.7}

\sethlcolor{boxcolorgray}

\def\gtrsim{\lower 2pt \hbox{$\, \buildrel {\scriptstyle >}\over
{\scriptstyle \sim}\,$}}
\def\lesssim{\lower 2pt \hbox{$\, \buildrel {\scriptstyle <}\over
{\scriptstyle \sim}\,$}}

\def\hst{{\sl HST}}
\def\jwst{{\sl JWST}}

\def\approxlt{\lower.2em\hbox{$\buildrel < \over \sim$}}
\def\approxgt{\lower.2em\hbox{$\buildrel > \over \sim$}}

\newcommand{\um}{\ensuremath{\mu\rm{m}}\xspace}

\newcommand{\Lir}{\ensuremath{L_{\rm{IR}}}\xspace}

\newcommand{\Msun}{\ensuremath{\rm{M}_\odot}\xspace}
\newcommand{\Lsun}{\ensuremath{\rm{L}_\odot}\xspace}

\newcommand{\Tdust}{\ensuremath{T_{\rm{dust}}}\xspace}
\newcommand{\Mdust}{\ensuremath{M_{\rm{dust}}}\xspace}

\newcommand{\Mstar}{\ensuremath{M_{\star}}\xspace}

\newcommand{\cii}{[C{\scriptsize II}]\xspace}

\def\be{\begin{equation}}
\def\ee{\end{equation}}
\def\bea{\begin{eqnarray*}}
\def\eea{\end{eqnarray*}}

\def\nii{[N{\sc ii}]}

\def\oiii{[O{\sc iii}]}

\def\jref@jnl#1{{\rm#1}}
\def\aj{\jref@jnl{AJ}}
\def\araa{\jref@jnl{ARA\&A}}
\def\apj{\jref@jnl{ApJ}}
\def\apjl{\jref@jnl{ApJ}}
\def\apjs{\jref@jnl{ApJS}}
\def\ao{\jref@jnl{Appl.~Opt.}}
\def\apss{\jref@jnl{Ap\&SS}}
\def\aap{\jref@jnl{A\&A}}
\def\aapr{\jref@jnl{A\&A~Rev.}}
\def\aaps{\jref@jnl{A\&AS}}
\def\azh{\jref@jnl{AZh}}
\def\baas{\jref@jnl{BAAS}}
\def\jrasc{\jref@jnl{JRASC}}
\def\memras{\jref@jnl{MmRAS}}
\def\mnras{\jref@jnl{MNRAS}}
\def\pra{\jref@jnl{Phys.~Rev.~A}}
\def\prb{\jref@jnl{Phys.~Rev.~B}}
\def\prc{\jref@jnl{Phys.~Rev.~C}}
\def\prd{\jref@jnl{Phys.~Rev.~D}}
\def\pre{\jref@jnl{Phys.~Rev.~E}}
\def\prl{\jref@jnl{Phys.~Rev.~Lett.}}
\def\pasp{\jref@jnl{PASP}}
\def\pasj{\jref@jnl{PASJ}}
\def\qjras{\jref@jnl{QJRAS}}
\def\skytel{\jref@jnl{S\&T}}
\def\solphys{\jref@jnl{Sol.~Phys.}}
\def\sovast{\jref@jnl{Soviet~Ast.}}
\def\ssr{\jref@jnl{Space~Sci.~Rev.}}
\def\zap{\jref@jnl{ZAp}}
\def\nat{\jref@jnl{Nature}}
\def\galax{\jref@jnl{Galaxies}}
\def\natas{\jref@jnl{Nat.~Astron.}}
\def\ore{\jref@jnl{Res.~Notes~AAS}}
\def\ittst{\jref@jnl{IEEE~Trans.~THz~Sci.~Technol.}}
\def\iaucirc{\jref@jnl{IAU~Circ.}}
\def\aplett{\jref@jnl{Astrophys.~Lett.}}
\def\apspr{\jref@jnl{Astrophys.~Space~Phys.~Res.}}
\def\bain{\jref@jnl{Bull.~Astron.~Inst.~Netherlands}}
\def\fcp{\jref@jnl{Fund.~Cosmic~Phys.}}
\def\gca{\jref@jnl{Geochim.~Cosmochim.~Acta}}
\def\grl{\jref@jnl{Geophys.~Res.~Lett.}}
\def\jcp{\jref@jnl{J.~Chem.~Phys.}}
\def\jgr{\jref@jnl{J.~Geophys.~Res.}}
\def\jqsrt{\jref@jnl{J.~Quant.~Spec.~Radiat.~Transf.}}
\def\memsai{\jref@jnl{Mem.~Soc.~Astron.~Italiana}}
\def\nphysa{\jref@jnl{Nucl.~Phys.~A}}
\def\physrep{\jref@jnl{Phys.~Rep.}}
\def\physscr{\jref@jnl{Phys.~Scr}}
\def\planss{\jref@jnl{Planet.~Space~Sci.}}
\def\procspie{\jref@jnl{Proc.~SPIE}}

\def\farcs{\hbox{$.\!\!^{\prime\prime}$}}

\def\ion#1#2{#1$\;${\small\rm\@Roman{#2}}\relax}

\newcount\lecurrentfam
\def\LaTeX{\lecurrentfam=\the\fam \leavevmode L\raise.42ex
\hbox{$\fam\lecurrentfam\scriptstyle\kern-.3em A$}\kern-.15em\TeX}
\def\sizrpt{
(\fontname\the\font): em=\the\fontdimen6\font, ex=\the\fontdimen5\font
\typeout{
(\fontname\the\font): em=\the\fontdimen6\font, ex=\the\fontdimen5\font
}}
\definecolor{dark}{rgb}{ 0.34, 0.09, 0.49}

\usepackage{fancyhdr,url}
\usepackage{transparent}
\usepackage{tablefootnote}
\usepackage{mathpazo}

\graphicspath{{figures/}}
\usepackage{booktabs}
\usepackage{titlesec}
\titlespacing*{\section}{0pt}{9pt plus 2pt minus 2pt}{4pt plus 1pt minus 1pt}
\titlespacing*{\subsection}{0pt}{7pt plus 2pt minus 2pt}{3pt plus 1pt minus 1pt}
\titlespacing*{\paragraph}{0pt}{3pt plus 1pt minus 1pt}{0.8em}

\newcommand{\aff}[2]{\hyperlink{#1}{\textsuperscript{#2}}}
\newcommand{\affitem}[2]{%
  \hypertarget{#1}{}#2\par
}

\begin{document}

\noindent
\begin{raggedright}
{\Huge\bfseries Mapping the Spatially Resolved Dust and Star Formation Physics Driving High-$z$ Galaxy Assembly\\[0.5em] \Large Science Prospects with ALMA 2040
}\end{raggedright}\\

\noindent
Daizhong Liu\aff{aff1}{1},
Gerg\"{o} Popping\aff{aff2}{2}, %
Pierre Cox\aff{aff3}{3}, 
Andreas Faisst\aff{aff4}{4}, %
Mamta Pandey-Pommier\aff{aff5}{5}, 
Dieu Nguyen\aff{aff6}{6}, 
Rom\'{a}n Fernandez Aranda\aff{aff7}{7}\textsuperscript{,}\aff{aff8}{8}, %
Elisabete da Cunha\aff{aff9}{9}\textsuperscript{,}\aff{aff10}{10}, %
Andrea Lapi\aff{aff11}{11}, 
Tanio D\'{i}az Santos\aff{aff8}{8}\textsuperscript{,}\aff{aff12}{12}, %
Ryota Ikeda\aff{aff13}{13}, 
Jacqueline Hodge\aff{aff13}{13}, %
Isabella Lamperti\aff{aff14}{14}, 
Manuel Aravena\aff{aff15}{15}\textsuperscript{,}\aff{aff16}{16}, 
Irene Shivaei\aff{aff14}{14}, 
Adarsh Kumar\aff{aff17}{17}, 
Roberto Decarli\aff{aff18}{18}, 
Mahmoud Hamed\aff{aff14}{14}, 
Umberto Maio\aff{aff19}{19}\textsuperscript{,}\aff{aff20}{20}, 
Gareth Jones\aff{aff21}{21}\textsuperscript{,}\aff{aff22}{22}, 
Jianhang Chen\aff{aff23}{23}, 
Sylvia Adscheid\aff{aff24}{24}
\footnote[1]{Affiliations are attached in the extra page.}

\begin{tcolorbox}[boxsep=3pt,left=4pt,right=4pt,top=3pt,bottom=3pt,colframe=boxcolor,colback=boxcolor]
\textbf{Abstract:} Interstellar dust reprocesses approximately half of the starlight emitted across cosmic history. However, dust temperature, which is essential to convert continuum flux into dust mass, molecular gas mass, and obscured star-formation rates, remains largely unconstrained on sub-galactic scales at high redshift. While single-band ALMA imaging has characterized continuum sizes for hundreds of $z>1$ galaxies, single-frequency measurements cannot separate the dust mass column from the local dust heating. To date, spatially resolved ($\lesssim 0.5''$) multi-band dust temperatures exist in the literature for only $\approx10$ extreme systems (quasars, hyperluminous starbursts, or strongly lensed galaxies) at $z>1$, with ongoing programs within a few years adding at most a few dozen submillimeter or UV-bright galaxies. This still leaves typical main-sequence disks largely unexplored across cosmic history. With an anticipated $\gtrsim 30\times$ improvement in continuum sensitivity (compared to the current ALMA),
simultaneous multi-band receivers, and (sub-)kpc angular resolution,
ALMA2040 will transform resolved dust thermometry into a systematic survey science. Here, we present the scientific rationale and technical requirements, highlighting a future statistically meaningful survey with $\gtrsim 100$ representative main-sequence galaxies across cosmic epochs.
Realistic visibility simulations demonstrate that ALMA2040 measures radial temperature gradients with ${\sim}0.5$--$4$~K precision ($L_{\rm IR} \approx 10^{12}\,L_\odot$ in 2~h on-source at $z\approx2$). These measurements will break the degeneracy between intrinsic compactness and temperature gradients, deliver the first extinction-free, temperature-calibrated resolved star-formation relations, and uniquely trace radiative feedback and disk maturation from cosmic noon to cosmic dawn.
\end{tcolorbox}

\vspace{0.5em}

\section{Scientific context and open questions}

\subsection{The unconstrained dust temperature field at high redshift}
Approximately half of the star formation during the epoch of cosmic noon ($z\approx1$--$3$) is obscured by interstellar dust \citep{Madau2014}. Over the past decade, ALMA continuum imaging has fundamentally advanced our view of this obscured activity, revealing dust distributions on scales down to hundreds of parsecs (for a review, see \citealp{Hodge2020}). However, the monochromatic submillimeter (submm) continuum flux density, $S_\nu \propto M_{\rm dust}\,\kappa_\nu\,B_\nu(T_{\rm dust})$, is intrinsically degenerate between dust mass surface density $\Sigma_{\rm dust}$ and the local dust temperature field $T_{\rm dust}$. At high redshift, dust temperatures are almost universally assumed rather than measured: continuum-based molecular gas mass calibrations \citep{Scoville2016,LiuDaizhong2019b,Tacconi2020} and far-infrared star-formation rates \citep{Kennicutt2012} rely on fiducial temperatures of $T_{\rm dust}\sim25$--$35$~K drawn from integrated, unresolved spectral energy distribution (SED) templates of brighter samples \citep{CaseyCaitlinM2012,MagnelliB2012a,Magnelli2014}.

The consequences of this assumption are profound. On the Rayleigh--Jeans tail, inferred gas and dust masses, who are constrained to each other via the gas-to-dust ratio, scale inversely with temperature ($M_{\rm gas \ or \ dust} \propto S_\nu / T_{\rm dust}$; \citealp{Scoville2016}). Applying an isothermal assumption across a galaxy with a realistic negative temperature gradient (e.g., $40$~K in the nucleus falling to $25$~K in the outer disk) overestimates the central dust column by $\sim60\%$ while underestimating the outer disk by $\sim30\%$. This artificially steepens the inferred radial gas profile by a factor of $\approx2$, directly biasing the resolved Kennicutt-Schmidt relation and corrupting dynamical stability measurements such as the Toomre $Q$ parameter. Radiation-hydrodynamical simulations demonstrate that dust temperatures in high-$z$ galaxies are highly structured and subject to optical depth effects, such that single-band fluxes combined with global template temperatures can introduce substantial systematic errors in inferred star-formation rates and gas distributions \citep{Behrens2018,DiMascia2023}.

Recovering both the dust column density and the temperature field requires at least two well-separated, high-signal-to-noise continuum maps on matched physical scales (three if breaking the $T_{\mathrm{dust}}$-$\beta$ degeneracy). Meeting this requirement has remained prohibitive with the current ALMA array (see Sect.~\ref{sec:state}), restricting resolved temperature studies to isolated extreme systems (Sect.~\ref{sec:scarce}). ALMA2040 is specifically designed to overcome this bottleneck \citep{Tan2026,Hodge2026,Facchini2025}.

\subsection{Key science goals for ALMA2040}
\label{sec:goals}
We frame the scientific motivation around five core science goals (SG1--SG5):

\paragraph{SG1: Radial temperature profiles and disk equilibrium.}
What is the intrinsic radial dust-temperature profile of normal star-forming galaxies at cosmic noon? Local spirals like M31 exhibit mild negative gradients of $\approx5$--$10$~K from center to outskirts \citep{Bothun1992,Kennicutt2011,Galametz2012,Aniano2012,Hunt2015}; see also \citet{DraineBT2007b}, \citet{Draine2011} and \citet{Smith2012, Smith2021}. In contrast, the few mapped $z>4$ extreme systems, such as hyperluminous quasar hosts, display central-to-outer temperature contrasts of $\approx20$~K driven by nuclear heating \citep{Tsukui2023,FernandezAranda2025}. Measuring whether smooth gradients or localized thermal spikes are generic in normal galaxies is essential to understanding radiative transfer and ISM energy balance during the peak of cosmic galaxy assembly.

\paragraph{SG2: Disentangling morphological compactness from temperature gradients.}
A cornerstone discovery of ALMA is that dusty star formation at $z\approx2$ is concentrated within compact submm cores whose effective radii are a factor of $\approx2$ smaller than the optical and near-infrared stellar light \citep{SimpsonJM2015b,Hodge2016,Lang2019,TadakiKenichi2020b,LiJuno2024}. However, because continuum brightness is temperature-weighted, warmer dust in the galactic nucleus naturally produces an artificially compact apparent size at shorter wavelengths, as observed directly in lensed systems \citep{Spilker2016} and predicted by forward radiative transfer modeling \citep{Cochrane2019}. Spatially resolved temperature maps are required to decouple temperature gradients from mass column profiles, determining the true physical mass-size relation of high-$z$ dust disks.

\paragraph{SG3: The extinction-free, temperature-calibrated resolved star-formation law.}
At $z\approx1$--$2$, optical and near-infrared integral-field spectroscopy has mapped the resolved star-forming main sequence and Kennicutt-Schmidt relations on (sub-)kpc scales \citep{CanoDiaz2016,Hsieh2017,LiuQing2018,Ellison2020,Baker2022,Wuyts2013,NelsonEricaJune2016a}. 
Complementary ALMA observations of mid-$J$ CO lines have resolved Kennicutt-Schmidt relation in small samples at $z\sim1\text{--}3$ \citep{ChenCC2017,Ikeda2022,ArriagadaNeira2025,Westoby2026}. 
At $z\gtrsim5$, initial resolved relations have been constructed using \cii\ \citep{Vallini2023,LiJuno2024,Accard2025}, but line-based gas tracers are sensitive to complex local conditions including metallicity, ionization, excitation and dynamics (e.g., \citealt{Casavecchia2024, Casavecchia2025}). Dust emission provides critical independent constraints on these physical conditions. ALMA2040 will deliver the first extinction-free, temperature-calibrated resolved star-formation law at $z>2$, providing both the obscured SFR surface density ($\Sigma_{\rm SFR}$) and the cross-calibrated gas surface density ($\Sigma_{\rm gas}$) from a large sample of dust continuum thermometry. ALMA2040 is irreplaceable for unveiling these properties for hundreds of star-forming galaxies at $z>2$ which will have \jwst{} spatially-resolved strong-line metallicity from intergral field unit (IFU), grism reconstruction, and multi-shutter array slit-stepping observations (e.g., \citealt{WangX2022, JuM2025, LeeL2026}).

\paragraph{SG4: Dust, gas, and stellar geometry.}
How is dust distributed relative to stellar populations on (sub-)kpc scales (e.g., in a uniform screen, mixed throughout the disk, or localized around young birth clouds; \citealp{Charlot2000})? While Balmer decrement profiles at $z\approx1.4$ suggest central attenuation peaks \citep{NelsonEricaJune2016b} and resolved ALMA-\hst/\jwst{} comparisons reveal compact dust within extended stellar disks \citep{Lang2019}, spatial offsets between dust and molecular gas challenge simple energy-balance assumptions \citep[e.g.,][]{RiveraGabrielaCalistro2018}. Resolving dust temperatures provides the missing thermal dimension necessary to place strong constraints on the dust-star geometries. Spatially-resolved SED fitting also breaks age/dust degeneracies \citep{LiJuno2024}. To achieve this goal, a large sample of high-$z$ disks with different geometries is needed. However, mapping their dust emission far exceeds the current ALMA continuum capability.

\paragraph{SG5: Radiative heating sources and feedback mechanisms.}
What fractions of dust heating at high redshift arise from young massive stars, diffuse interstellar radiation from older populations (as seen in M31's bulge; \citealt{Groves2012,Viaene2014}), AGN radiation, or mechanical dissipation? Spatially resolved temperature maps down to (sub-)kpc scales to resolve the nuclear region and galactic sub-structures would provide the breakthrough diagnostic, distinguishing compact nuclear hot spots powered by AGN accretion \citep{Tsukui2023,MeyerRomainA2025a} from widespread mechanical heating across tens of kpcs \citep{FernandezAranda2025} at $z > 2$.

\subsection{The state of the art: extensive continuum sizes, assumed temperatures}
\label{sec:state}
High-resolution single-band surveys have mapped obs.-frame 870\,$\mu\mathrm{m}$ dust structures for hundreds of SMGs and typical main-sequence galaxies at $z\approx2\text{--}6$ (e.g., \citealp{SimpsonJM2015b,Hodge2016,Hodge2019,Dudzevivciute2020,Smail2021,LiuDaizhong2019a,Adscheid2024,TadakiKenichi2020b,Lang2019,LeFevre2020,Ikeda2022,HerreraCamus2025,Pozzi2024,MitsuhashiIkki2024a,LiJuno2026}), typically finding compact effective radii ($r_{50}\sim0.5\text{--}2.5$\,kpc), systematically a factor of $\approx2$ smaller than their stellar light \citep[see theoretical discussion in][]{Popping2022}. However, single-band emission is biased to a temperature-weighting rather than tracing intrinsic mass column, and the warmer central regions appear more compact at shorter submm wavelengths near the SED peak \citep{Spilker2016,Cochrane2019,Behrens2018}.

\begin{figure}[tbp]
\centering
\begin{minipage}{0.88\textwidth}
\centering
{\footnotesize (i) simple exponential disk ($z=2$), $M_{\rm d}=10^{8.5}\,\Msun$,  $L_{\rm IR}\simeq10^{12.0}\,\Lsun$ (2\,h)}\\[-0.2ex]
\includegraphics[width=\textwidth]{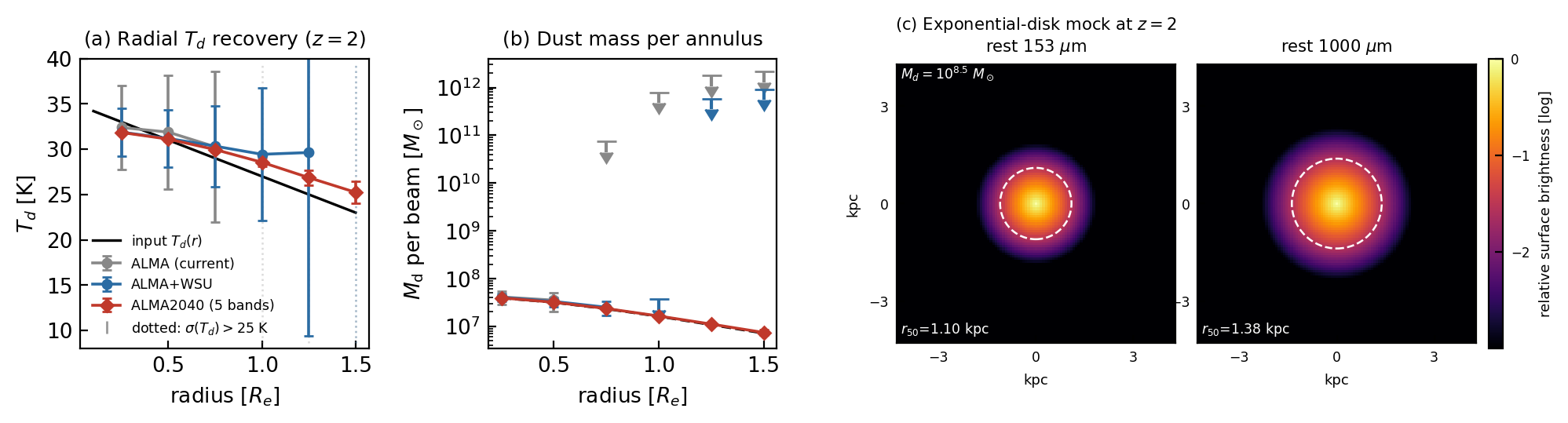}\\
{\footnotesize (ii) NGC4321 mock ($z=2$), $M_{\rm d}=10^{8.5}\,\Msun$, $L_{\rm IR}\simeq10^{12.0}\,\Lsun$ (2\,h)}\\[-0.2ex]
\includegraphics[width=\textwidth]{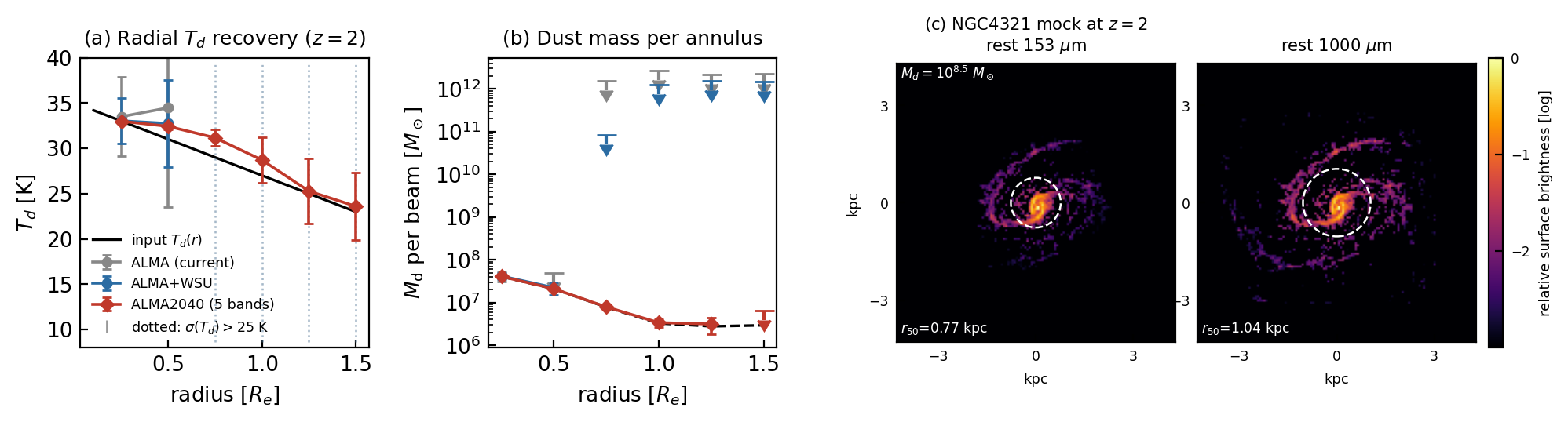}\\
{\footnotesize (iii) NGC4321 mock ($z=5$), $M_{\rm d}=10^{8.5}\,\Msun$, $L_{\rm IR}\simeq10^{12.0}\,\Lsun$ (8\,h, main sequence)}\\[-0.2ex]
\includegraphics[width=\textwidth]{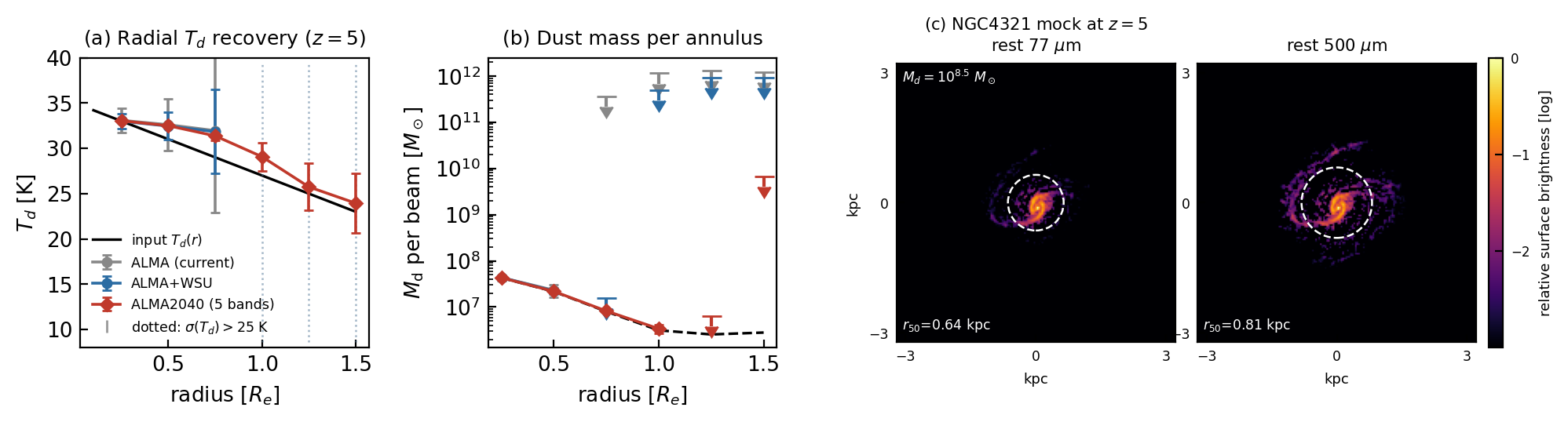}\\
\end{minipage}
\vspace{-1.7ex}
\caption{Mock galaxy feasibility tests for representative $L^\star$ star-forming galaxies ($M_{\rm dust}\approx 3\times 10^8\,\Msun$, $L_{\rm IR}\simeq 10^{12}\,\Lsun$):
\textbf{(a)} Radial dust temperature recovery $T_{\rm d}(r)$ in 1.5-kpc synthesized beams comparing current ALMA (grey circles), ALMA+WSU (blue circles), and ALMA2040 5-band simultaneous observations (red diamonds), for 2\,h on source (rows~i and~ii) and 8\,h on source (row~iii). Dotted vertical lines mark annuli where fits become unconstrained ($\sigma(\Tdust) > 25$\,K).
\textbf{(b)} Precision of dust mass per beam constraint in each annulus. While current ALMA and WSU fail beyond $r\approx 0.5\text{--}0.75\,r_{50}$ (producing unconstrained upper limits), ALMA2040 achieves $0.6\text{--}1\%$ $T_{\rm dust}$ accuracy and $1\text{--}2\%$ $M_{\rm dust}$ accuracy at $0.5\,r_{50}$ across $z\sim 2\text{--}5$, tracing the intrinsic mass profile out to $1.5\,r_{50}$.
\textbf{(c)} Model surface brightness maps at short and long rest-frame submm wavelengths (dashed circles mark apparent $r_{50}$), demonstrating artificial wavelength-dependent compaction driven by radial temperature gradients.
}
\label{fig:recovery}
\end{figure}

This effect can be quantified using a fiducial exponential disk ($r_{50} = 1.8$\,kpc at $z=2$) with a modest radial temperature gradient of $8$\,K per $r_{50}$ ($\Lir=10^{12}\,\Lsun$, $\Mdust\approx3 \times 10^8\,\Msun$; Figure~\ref{fig:recovery}, top row). The apparent flux-weighted half-light radius shifts from $1.38$\,kpc at rest $1000\,\mu\mathrm{m}$ to $1.10$\,kpc at rest $153\,\mu\mathrm{m}$ (a $25\%$ apparent size change across bands), while the true dust mass profile radius is $1.8$\,kpc. Furthermore, applying an isothermal 30\,K conversion overestimates the central dust column by $+40\%$ and underestimates the column at $r_{50}$ by $-31\%$ (Figure~\ref{fig:recovery}; cf. \citealp{DiMascia2023}). Decoupling genuine morphological compaction from radial temperature profiles requires spatially resolved multi-band thermometry.

\subsection{Current scarcity of resolved temperature measurements}
\label{sec:scarce}
\begin{figure}[t]
\centering
\includegraphics[width=0.78\textwidth]{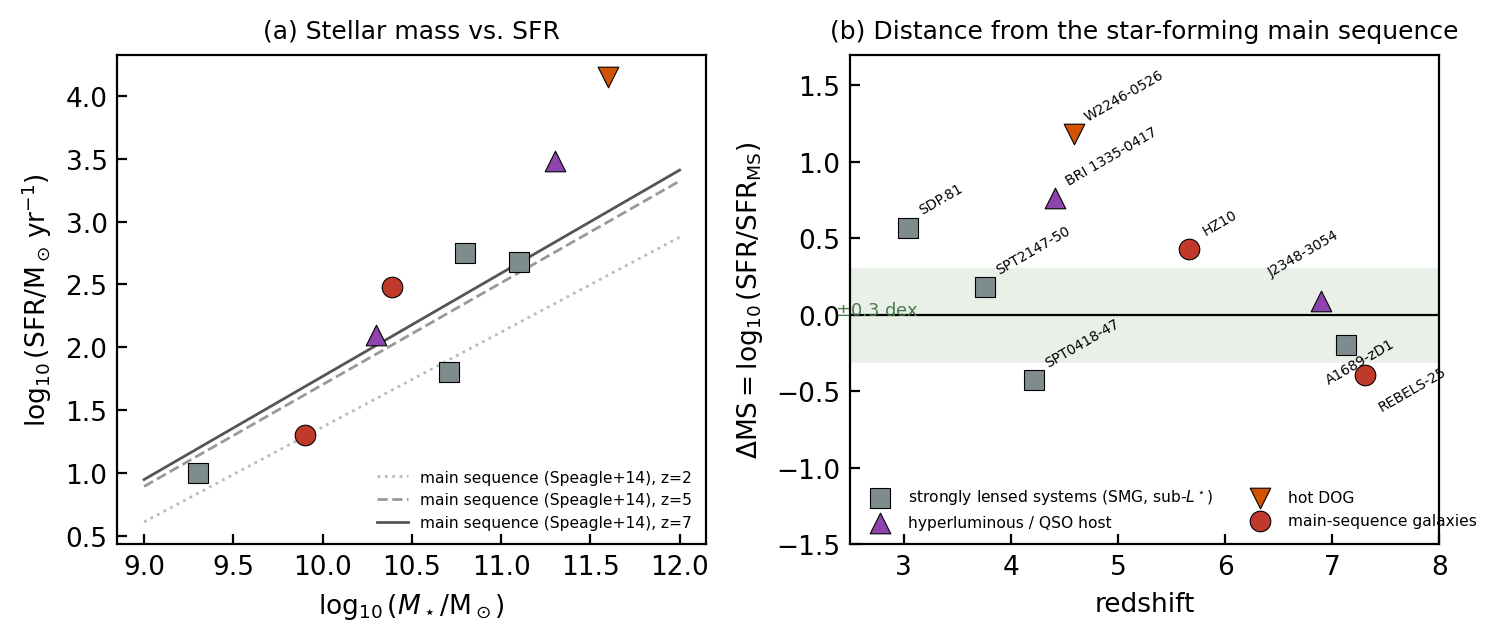}
\vspace{-1.7ex}
\caption{The Table~\ref{tab:census} sample in the physical plane. (a) $M_{\star}$ vs.
SFR; lines show the star-forming main sequence of \citet{Speagle2014} at $z=2$, 5, and 7. (b) Offset from the main sequence, $\Delta\mathrm{MS}=\log_{10}(\mathrm{SFR}/\mathrm{SFR}_{\rm MS})$, vs. $z$; the shaded band marks $\pm0.3$~dex. (See references in Table~\ref{tab:census}).}
\label{fig:scarcity}
\end{figure}

\begin{figure}[t]
\centering
\includegraphics[width=0.96\textwidth]{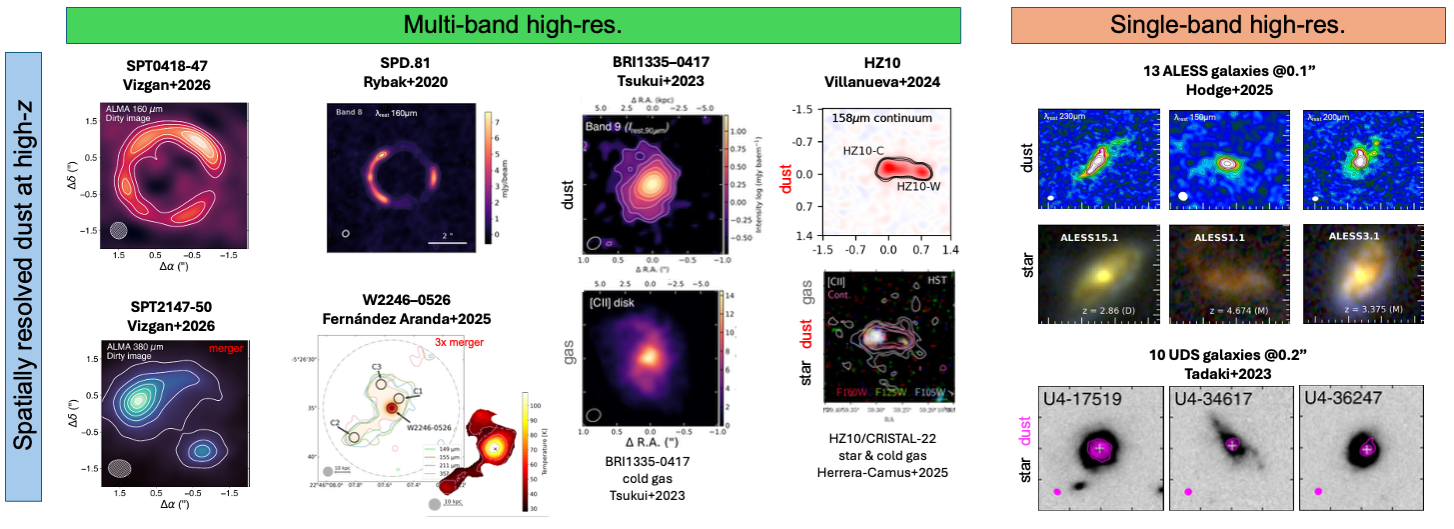}
\vspace{-1.7ex}
\caption{Resolved dust-continuum maps delivered by current ALMA at high redshift, compiled from the literature. Multi-band constraints (green) are shown for six systems as best examples, which are dominated by quasar hosts, hyperluminous starbursts, lensed systems, and a previously-thought $z \sim 5$ main-sequence galaxy, HZ10, which turned out to be a merger of three galaxies (\citealt{Telikova2025, Jones2025}). Single-band imaging (orange) achieves higher fidelity, but disk and spiral morphology remain rare and temperature-column degeneracies persist.}
\label{fig:sample}
\end{figure}

Measuring resolved dust temperatures requires at least two high-S/N continuum bands (and ideally multiple to solve both $\beta$ and $T_{\mathrm{d}}$) on matched physical scales. At $z>1$, published resolved ($\lesssim 0.2\text{--}0.5''$) multi-band continuum data exist for only a dozen or so galaxies (Table~\ref{tab:census}, Figure~\ref{fig:scarcity}, and Figure~\ref{fig:sample}). These systems are strongly biased toward extreme objects: quasar hosts with central temperatures of $55\text{--}110$\,K driven by AGN heating (e.g., BRI\,1335$-$0417, W2246$-$0526, J2348$-$3054; \citealp{Tsukui2023,FernandezAranda2025,MeyerRomainA2025a}) or gravitationally lensed starbursts (e.g., SDP.81, SPT sources; \citealp{Rybak2015,Rybak2020,Vizgan2026}). Only one source approaches normal star-forming galaxies (SFGs): the strongly-lensed sub-$L^*$ galaxy A1689-zD1 at $z=7.13$ ($T_{\rm dust}=41^{+17}_{-14}$\,K; \citealp{Akins2022,Witstok2022,Vallini2023}), as even HZ10 (CRISTAL-22) at $z=5.66$, ($T_{\rm dust}\approx51\pm13$\,K; \citealt{Villanueva2024}) has recently been resolved into a chaotic triple merger \citep{Telikova2025,Jones2025}. Undisturbed main-sequence disks remain largely unexplored.

This census demonstrates that existing high-$z$ temperature maps are restricted to extreme regimes and deliver uncertainties of $\pm10\text{--}15$\,K per system, which is insufficient to resolve the subtle $5\text{--}10$\,K gradients typical of normal disks. An archive-scale analysis reinforces this limitation: in the A3COSMOS homogeneous archive covering the COSMOS field \citep{LiuDaizhong2019a, Adscheid2024}, $\approx280$ galaxies are detected in two or more ALMA bands, but only $\approx10$ possess multi-band data at high angular resolution ($<0\farcs5$). Resolved dust temperatures and star-formation physical properties of normal galaxies is fundamentally beyond the throughput of the current array.

\subsection{Why ALMA2040 is essential}
While the Wideband Sensitivity Upgrade (WSU) will increase continuum imaging speed by a factor of $\sim3$ \citep{Carpenter2019,Carpenter2023}, bridging the gap from isolated case studies to systematic surveys of typical galaxies requires ALMA2040 \citep{Tan2026,Hodge2026,Facchini2025}. With an expanded collecting area, wider instantaneous bandwidth, and simultaneous multi-band receivers, ALMA2040 delivers a \textit{transformational} improvement in continuum sensitivity (Sect.~2.1), enabling radial temperature profiling at $\lesssim 1$\,K precision across representative galaxy populations.

\begin{table}[t]
\footnotesize
\centering
\caption{%
\footnotesize
The published sample of highly spatially resolved ($\lesssim 0.2''$) multi-band dust-continuum constraints at
$z>1$ (selected cases to our knowledge, mid-2026; likely incomplete), together with $M_{\star}$ and SFRs compiled for Figure~\ref{fig:scarcity}. This means two or more continuum bands imaged and analyzed at comparable (sub-kpc to kpc) resolution. Columns~5 and~6 give $\log_{10}(M_\star/\Msun)$ and $\log_{10}(\mathrm{SFR}/\Msun\,\mathrm{yr}^{-1})$; all values are intrinsic physical properties, i.e. delensed (corrected for lensing magnification) for the lensed sources. SFRs of the AGN-dominated systems (BRI~1335$-$0417, W2246$-$0526, J2348$-$3054) are effectively upper limits. \emph{References:} (1)~\citet{Rybak2015}; (2)~\citet{Rybak2020}; (3)~\citet{Vizgan2026}; (4)~\citet{Tadaki2018}; (5)~\citet{Ikeda2026a}; (6)~\citet{Tsukui2023}; (7)~\citet{FernandezAranda2025}; (8)~\citet{Villanueva2024}; (9)~\citet{MeyerRomainA2025a}; (10)~\citet{Akins2022}; (11)~\citet{Algera2024}.}
\label{tab:census}
\vspace{-4ex}
\begin{tabular}{@{}l c l l c c c@{}}
\hline\hline
Source & $z$ & Type & Rest $\lambda$\,[$\um$] & $\log(M_\star/\Msun)$ & $\log\mathrm{SFR}$ & Ref. \\
\hline
SDP.81 & 3.04 & lensed SMG & 220, 270 & 10.8 & 2.75 & 1, 2 \\
SPT2147$-$50 & 3.76 & lensed SMG & 160, 380 & 11.1 & 2.68 & 3 \\
SPT0418$-$47 & 4.22 & lensed SMG & 160, 380 & 10.71 & 1.80 & 3 \\
AzTEC-1 & 4.34 & SMG & 160, 240, 380, 660 & 10.91 & 2.86 & 4,5 \\
BRI 1335$-$0417 & 4.41 & hyper-lum. QSO & 90, 161 & 11.3 & 3.48 & 6 \\
W2246$-$0526 & 4.59 & hot DOG & 60--600 (B3--10) & 11.6 & 4.15 & 7 \\
HZ10 (CRISTAL-22) & 5.66 & main-sequence & 77, 130 & 10.39 & 2.48 & 8 \\
J2348$-$3054 & 6.90 & QSO host & 160, 350 & 10.3 & 2.10 & 9 \\
A1689-zD1 & 7.13 & lensed sub-$L^\star$ & 52--400 & 9.3 & 1.00 & 10 \\
REBELS-25 & 7.31 & LBG, massive disk & multi & 9.9 & 1.30 & 11 \\
\hline
\end{tabular}
\vspace{-1ex}
\end{table}

\section{ALMA2040: Technical Requirements}

\subsection{Observational capabilities, receiver architecture, and array configurations}
\label{sec:tech}
Measuring spatially resolved dust temperatures is fundamentally limited by continuum surface brightness sensitivity. Resolving (sub-)kpc scales across high-$z$ galaxy disks requires three coordinated capabilities under the ALMA2040 vision.

\begin{figure}[t]
\centering
\includegraphics[width=0.72\textwidth]{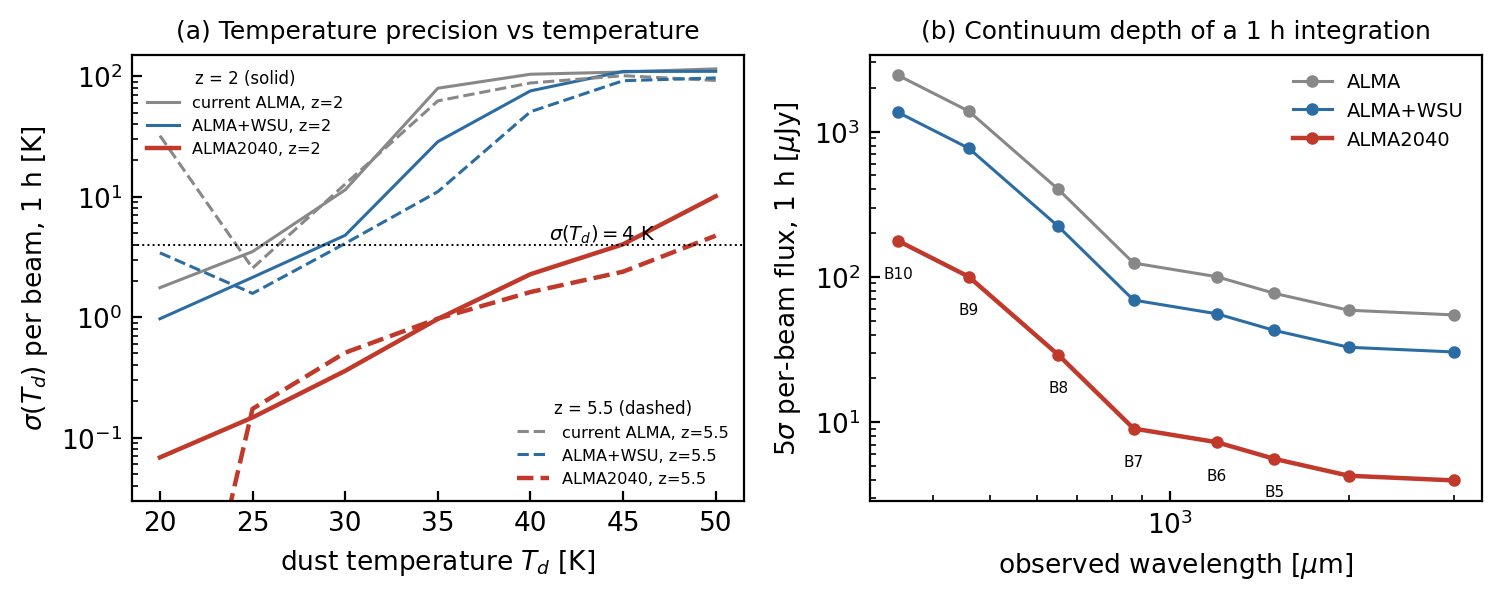}
\vspace{-1.7ex}
\caption{%
Monte Carlo dust temperature precision per 1.5-kpc synthesized beam in 1~h on source as a function of $T_{\rm dust}$, comparing current ALMA (grey), ALMA+WSU (blue), and ALMA2040 (red; 255 antennas, 5-band simultaneously with 64~GHz IF bandwidth per band) for $z \sim 2$ (solid) and $z \sim 5$ (dashed). The horizontal dotted line marks $\sigma(T_{\rm d})=4$~K. While current ALMA and WSU degrade to $\sigma(T_{\rm d})>20\text{--}100$~K above $T_{\rm dust}\sim 30\text{--}35$~K, ALMA2040 maintains sub-Kelvin to few-Kelvin precision across the full range of typical interstellar dust temperature ($T_{\rm dust}\sim 20\text{--}45$~K).
\textbf{(b)} 5$\sigma$ continuum depth in 1~h across bands (Bands 5--10 labeled), demonstrating the $\sim 13\times$ per-band sensitivity gain achieved by ALMA2040.
}
\label{fig:bandpairs}
\end{figure}

\paragraph{(i) Continuum sensitivity scaling.}
The ALMA2040 design envisions an expanded collecting area (i.e., 200--300 antennas in extended configurations of $\sim 20$~km baseline), much wider instantaneous bandwidth (e.g., 512 GHz; here we assume 5-band multiplex, each with 64~GHz IF bandwidth), and upgraded quantum-limited receiver technology \citep{Tan2026,Baryshev2024}. This delivers a factor of $\sim 13 \times$ gain in point-source continuum sensitivity per band, and an aggregate gain of $\sim 30 \times$ across the 5 bands, boosting survey speed by $\sim 900 \times$ to reach a given continuum depth, relative to the current array (Figure~\ref{fig:bandpairs}b).
Because temperature uncertainty from a multi-band modified blackbody fit scales inversely with the signal-to-noise ratio in the faintest band, this sensitivity gain improves temperature precision $\sim 10 \text{--} 30 \times$ at fixed exposure. For a fiducial $z=2$ main-sequence disk with a peak $T_{\mathrm{dust}}\approx30$\,K, a 1-h integration yields $\sigma(\Tdust)\approx11$\,K with the current ALMA, $\approx5$\,K with WSU, and $\approx 0.35$\,K with the 5-band ALMA2040 assumption (Figure~\ref{fig:bandpairs}a), maintaining $\sigma(\Tdust)\lesssim 1\text{--}4$\,K across typical disk temperatures ($T_{\rm dust}\sim 20\text{--}45$\,K) where current ALMA becomes completely unconstrained ($\sigma(\Tdust)\gg 50$\,K).
ALMA2040 is the only facility that can extend radial temperature and dust mass profile recovery out to $\sim 1 \text{--} 1.5\,r_{50}$ at $z\sim 2 \text{--} 5$ even with complex, realistic dust distributions, delivering $0.5\text{--}1\%$ precision in $T_{\rm dust}$ and $1\text{--}2\%$ precision in $M_{\rm dust}$ at $0.5\,r_{50}$ (Figure~\ref{fig:recovery}; $\sim 30\times$ better than the current ALMA).

\paragraph{(ii) Simultaneous multi-band observing and atmospheric calibration.}
Simultaneous multi-band receivers represent a qualitative transformation for continuum thermometry \citep{Tan2026}. Multi-band optics utilizing frequency-selective surfaces (dichroic beam splitters) and diplexers are currently demonstrating simultaneous coverage for Bands~6+7 \citep{Masui2020}, Bands~7+8 \citep{Kojima2018}, and Bands~6+9 \citep{Montofre2017}. Future ngEHT telescopes and the Chinese XSMT-15m telescope (\citealt{XSMT2025}) will all be equipped with 3-band heterodyne receivers.
Observing $\ge 2$ continuum bands simultaneously through the identical atmosphere, telescope optics, and baseline geometry theoretically cancels out time-variable atmospheric fluctuations, pointing drift, and relative flux calibration uncertainties that currently make sequential multi-band analyses much nosier. Furthermore, simultaneous multi-band capability enables \emph{Band-to-Band (B2B) phase transfer} (\citealt{Issaoun2025}): rapid atmospheric phase fluctuations measured continuously in the lower-frequency, high-coherence band (e.g., Band~7) can be directly applied to stabilize the high-frequency band (Band~9). This substantially extends coherence times, relaxes calibrator choices, and maximizes imaging fidelity during submm observations.

At the Chajnantor plateau, high-frequency operations are governed by atmospheric precipitable water vapor (PWV).
While prime Octile 1 weather ($\mathrm{PWV}<0.5$\,mm, $\sim 10$--15\% availability) enables the full five-band suite (Bands 6--10), the simultaneous receiver architecture provides natural operational resilience: in Octile 2--3 conditions ($\mathrm{PWV} \approx0.6$--1.2\,mm, $\sim 40\%$ availability), Bands 6+7+8 (or Bands 4--8) still observe simultaneously, delivering high-fidelity multi-band thermometry without wasting observing windows.

\paragraph{(iii) Angular resolution and spatial filtering considerations.}
Resolving $1\text{--}2$\,kpc physical scales at $z\approx2\text{--}5$ requires synthesized beam sizes of $0\farcs12\text{--}0\farcs25$. In ALMA2040, these scales are achievable by combining Compact (0.4\,km baseline) + Intermediate (12.5\,km baseline) configurations, with Maximum Recoverable Scales $\mathrm{MRS} \approx 1\farcs5\text{--}5\farcs4$ across Bands~6--10. This combination preserves the total flux of extended high-$z$ galactic disks while delivering high surface-brightness sensitivity.
Furthermore, the ALMA2040 Extended configuration (synthesized beam $\approx0\farcs0087$ in Band~7, $\mathrm{MRS} = 0\farcs194$) would spatially resolve the high-$z$ nuclear region, extremely powerful in imaging high-$T_{\mathrm{dust}}$ starburst cores or clumps.

\begin{figure}[t]
\centering
\includegraphics[width=0.78\textwidth]{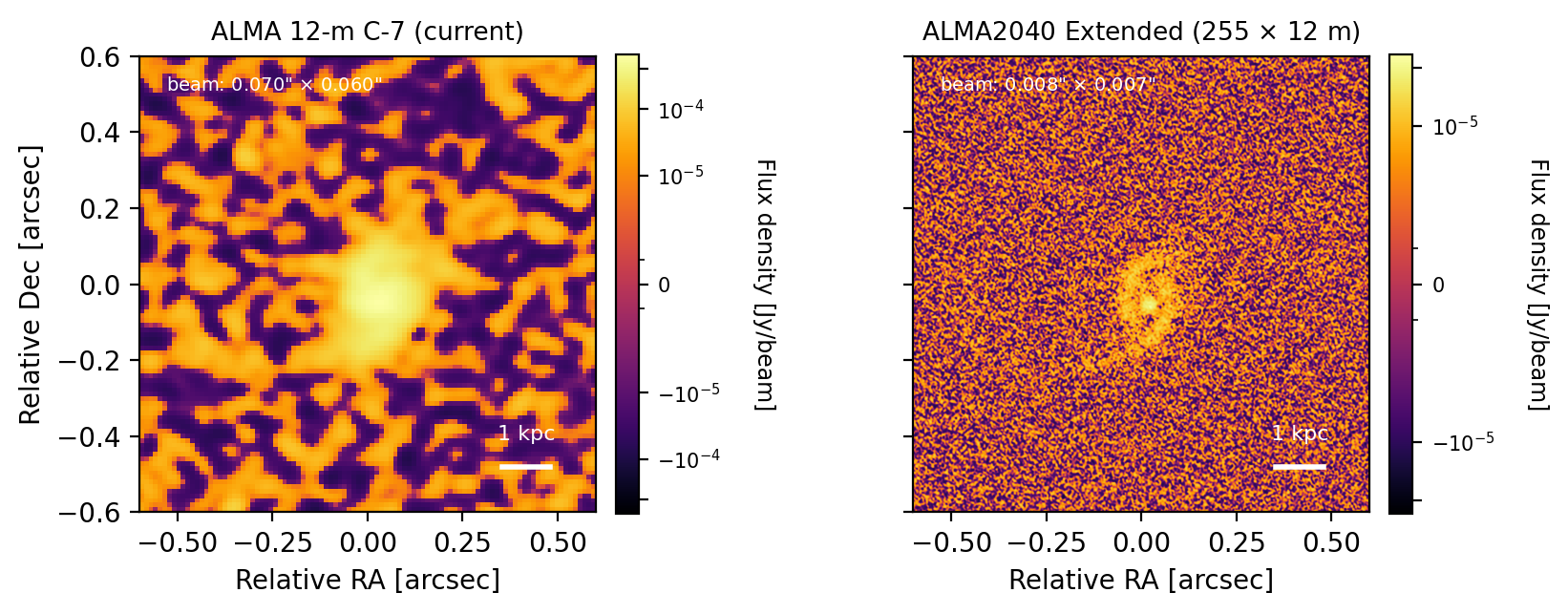}
\vspace{-1.7ex}
\caption{End-to-end \textsc{Ghost} visibility simulations of the NGC4321-based $z=3$ sky model with
$M_{\rm d}=10^{8.5}\,\Msun$ (Band~7, $345$\,GHz, bandwidth $\times$  on-source integration time $\Delta \nu \times \Delta t$ = 7.5\,GHz$\cdot$h, Briggs robust 0.5): left: current ALMA C-7 configuration; right: ALMA2040 array (but with bandwidth of only 7.5~GHz as limited by the current \textsc{Ghost} version). ALMA2040 recovers the extended spiral disk with order-of-magnitude noise reduction.}
\label{fig:ghost}
\end{figure}

\paragraph{(iv) End-to-end visibility simulations with GHOST.}
To evaluate observational feasibility, we simulated the NGC4321-based $z=3$ sky model ($M_{\rm d}=10^{8.5}\,\Msun$; see also Figure~\ref{fig:recovery}) using \textsc{Ghost}, the official ALMA2040 simulation tool ({\footnotesize \url{https://alma-ghost.jb.man.ac.uk/}}). We observed Band~7 (center frequency $350$\,GHz, bandwidth $7.5$\,GHz\,\footnote[2]{The bandwidth is limited by the current version of GHOST which will be updated in the next stages.}) for 1\,h on source using the current ALMA 12-m C-7 configuration and the ALMA2040 array, applying Briggs weighting (robust 0.5) and CLEAN deconvolution (Figure~\ref{fig:ghost}). The simulation highlights the transformative imaging fidelity. Even in the extreme, highest-resolution ALMA2040 Extended configuration (beam $\sim 8$\,mas, or $\sim 60$~pc at $z\sim3$), ALMA2040 clearly recovers dense clump-scale spiral arms that are completely blurred into noise with current ALMA.

Two foundational design principles emerge from Figures~\ref{fig:bandpairs}--\ref{fig:ghost}:\vspace{-1.5ex}
\begin{enumerate}\setlength{\itemsep}{0pt}\setlength{\parskip}{0pt}
\item \textbf{Color leverage peaks in the submm transition regime:}
For typical cold dust ($T_{\rm dust}\approx30\text{--}40$\,K, $\beta\approx1.8\text{--}2.0$), the modified blackbody peak in $S_\nu$ occurs at $\lambda_{\rm rest}\approx75\text{--}95\,\mu\mathrm{m}$. At $z\approx2$, Bands~6--10 simultaneously cover $\lambda_{\mathrm{rest}}\approx115\text{--}430\,\mu\mathrm{m}$ sampling 5 photometric points spanning the Rayleigh-Jeans tail and pushing directly into the turnover near the peak.
At $z\approx5$, Bands~6--10 cover $\lambda_{\mathrm{rest}}\approx53\text{--}119\,\mu\mathrm{m}$, bracketing the dust SED peak.
This breaks the temperature-invariance of continuum color, yielding fractional temperature uncertainties $\sigma(\Tdust)/\Tdust$ of a few percent.
\item \textbf{Decoupling the $T_{\rm dust}\text{--}\beta$ degeneracy:}
With two continuum bands, $T_{\rm dust}$ is constrained under physically motivated priors on the emissivity index (e.g., $\beta=1.8\pm0.2$; \citealp{Algera2024}). Because spatial variations of $\Delta\beta\approx0.2$ produce apparent temperature shifts of $\approx4\text{--}5$\,K, observing $\ge 3$ bands simultaneously across the SED peak enables directly solving for $M_{\rm dust}$, $T_{\rm dust}$, and $\beta$ per resolution element.
\end{enumerate}

\subsection{A transformational survey: mapping dust and star-formation physics from cosmic dawn to noon}
These capabilities motivate a reference ALMA2040 survey of $\ge 100$ representative main-sequence galaxies spanning cosmic dawn ($z\approx4\text{--}6$) to cosmic noon ($z\approx1\text{--}2$). Drawing from mass- and SFR-complete \jwst{} and radio censuses (e.g., $\Mstar \sim 10^{10\text{--}12}\,\Msun$, $\mathrm{SFR}\sim 20\text{--}1000\,\Msun/\mathrm{yr}$), simultaneously covering Bands~6--10 ($\lambda_{\mathrm{rest}}\approx 115\text{--}450\,\mu\mathrm{m}$ at $z\sim2$; $\approx 50\text{--}200\,\mu\mathrm{m}$ at $z\sim6$) maps dust emission out to $\sim 1.5\,r_{50}$ at $1\text{--}2$\,kpc ($0\farcs12\text{--}0\farcs25$) resolution in $1\text{--}2$\,h per target (can reach sub-kpc with more weighting on long baselines for bright centers). With the current ALMA, reaching the same sensitivity in a single band would demand $\approx 170\text{--}340$\,h per target ($\gtrsim 1,000$\,h sequentially per target), making such a survey impossible within normal TAC allocations.

This survey directly achieves the goals of Sect.~\ref{sec:goals}: (1) delivering radial temperature profiles with $\sigma(\Tdust)\approx2\text{--}4$\,K per beam, measuring $5\text{--}10$\,K gradients at $2\text{--}5\,\sigma$ significance (SG1); (2) eliminating the $\sim25\%$ temperature-induced apparent size shift to derive intrinsic dust mass profiles $\Sigma_{\rm dust}(r)$ and physical sizes (SG2); (3) combining with \jwst{} stellar mass and metallicity maps to calibrate the resolved Kennicutt-Schmidt law and dust-to-gas conversions (SG3, SG4); and (4) resolving localized nuclear heating to distinguish AGN accretion from merger shocks and stellar feedback (SG5).

\paragraph{Observational caveats and pipeline mitigations:}
(i) \emph{Line contamination}: \cii\ $158\,\mu\mathrm{m}$ falls into Band~9 at $z\approx2$ and Bands~6--7 at $z\approx4\text{--}6$, while \oiii\ $88\,\mu\mathrm{m}$ and \nii\ $122\,\mu\mathrm{m}$ enter Bands~8--9; wide instantaneous bandwidth guarantees ample line-free continuum channels while providing simultaneous kinematic tracers. (ii) \emph{Very small grains (VSGs)}: At $z\gtrsim5$, Bands~9--10 reach $\lambda_{\mathrm{rest}}\sim 50\text{--}75\,\mu\mathrm{m}$, where stochastically heated VSGs contribute a smooth, featureless continuum; multi-wavelength SED modeling incorporating realistic grain-size distributions (e.g., THEMIS or Draine \& Li) will parameterize and isolate this component from the cold equilibrium dust. (iii) \emph{CMB suppression}: At $z>4$, CMB heating and background contrast ($T_{\rm CMB} = 2.73\,(1+z)$\,K) produce $\le15\%$ flux reduction for typical $\ge35$\,K dust, routinely corrected in modified blackbody fits (\citealt{daCunha2013}; while the coldest dust gradually becomes undetectable against the CMB).

\section{Synergies with current and future facilities}
\label{sec:syn}
Resolved dust temperatures acquire the full physical context when matched to panchromatic datasets and theoretical modeling:

\paragraph{Panchromatic counterparts (\jwst{} and ELT):}
\jwst{} NIRCam imaging resolves stellar structures at $z\approx2\text{--}6$, while NIRSpec and MIRI IFUs map resolved nebular line diagnostics \citep{FaisstAL2026,Tsujita2026,HerreraCamus2026,Donnan2026}. Combining \jwst{}'s view of unobscured stars and ionized gas with ALMA2040's obscured star formation and cold dust provides a complete mass assembly budget on sub-kpc scales, which the Extremely Large Telescope (ELT) will extend to $\approx50$\,pc scales \citep{HerreraCamus2026}.

\paragraph{Radio and FIR/submm facilities (SKA, ngVLA, ALMA spec. lines, PRIMA, AtLAST):}
Ultra-deep radio continuum with the Square Kilometer Array (SKA) and ngVLA traces free-free and synchrotron emission at sub-arcsecond resolution, delivering SFRs independent of dust attenuation and temperature \citep{Condon1992,Behiri2024}.
ALMA/ALMA2040 spectral line surveys will provide important gas tracer calibrations for the portion of brighter high-$z$ SFGs among the dust samples (see companion ALMA2040 whitepapers).
Complementary cooled FIR space telescopes (e.g., PRIMA) will map global spectrophotometry across the $50\text{--}250\,\mu\mathrm{m}$ rest-frame peak \citep{Glenn2025}, while wide-field submm single dishes (e.g., AtLAST) will anchor the total zero-spacing flux down to a few arcsec for ultimate large SMG samples \citep{vanKampen2024}.

\paragraph{Numerical simulations and radiative transfer:}
Interpreting resolved dust temperatures requires high-resolution radiation-hydrodynamic simulations coupled to dust radiative transfer \citep{Cochrane2019,Behrens2018,DiMascia2023,Orr2018,Vallini2023}. Community synthetic multi-band ALMA2040 visibility benchmarks will calibrate temperature-recovery pipelines, configurations, and $\beta$ priors ahead of operations.

\vspace{-2ex}
\section*{Acknowledgements}

\indent This work was developed as part of the broader ALMA2040 community effort. We acknowledge the Lorentz Center and NOVA for supporting the community workshops and coordination activities that helped initiate and shape the ALMA2040 White Paper process. 
We thank the ALMA2040 high-$z$ working group leads T. Bakx, F. Rizzo and H. Stacey for the great organizing work. We also thank the referees of the ALMA2040 internal review process for providing constructive comments and suggestions. 
DL acknowledges the support from the Strategic Priority Research Program of the Chinese Academy of Sciences, grant No. XDB0800401.

\bibliographystyle{apj_mod}
{\footnotesize
\bibliography{bibliography,manual_entries}}

\newpage

\section*{Supplementary Information}

\vspace{0.8em}

{\large\textbf{Author affiliations}}

\vspace{0.8em}
{
\small{
\noindent \affitem{aff1}{$^{1}$ State Key Laboratory of Radio Astronomy and Technology, Purple Mountain Observatory, Chinese Academy of Sciences, 10 Yuanhua Road, Nanjing 210023, China}
\noindent \affitem{aff2}{$^{2}$ European Southern Observatory, Karl-Schwarzschild-Str. 2, D-85748, Garching, Germany}
\noindent \affitem{aff3}{$^{3}$ Sorbonne Universit\'{e}, UPMC Paris 6 and CNRS, UMR 7095, Institut d'Astrophysique de Paris, 98b bd. Arago, 75014 Paris, France}
\noindent \affitem{aff4}{$^{4}$ Caltech/IPAC, 1200 E. California Boulevard, Pasadena, CA 91125, USA}
\noindent \affitem{aff5}{$^{5}$ {Pole Scientific, University Catholic of Lyon- University of Lyon, 10 place des Archives 69002, Lyon, France}}
\noindent \affitem{aff6}{$^{6}$ {Department of Astronomy, University of Michigan, 1085 South University Avenue, Ann Arbor, MI 48109, USA}}
\noindent \affitem{aff7}{$^{7}$ Department of Physics, University of Crete, 70013 Heraklion, Greece}
\noindent \affitem{aff8}{$^{8}$ Institute of Astrophysics, Foundation for Research and Technology - Hellas (FORTH), 70013 Heraklion, Greece}
\noindent \affitem{aff9}{$^{9}$ International Centre for Radio Astronomy Research (ICRAR), The  University of Western Australia, M468, 35 Stirling Highway, Crawley, WA 6009, Australia}
\noindent \affitem{aff10}{$^{10}$ ARC Center of Excellence for All Sky Astrophysics in 3 Dimensions (ASTRO 3D), Canberra, Australia}
\noindent \affitem{aff11}{$^{11}$ SISSA, Via Bonomea 265, 34136 Trieste, Italy}
\noindent \affitem{aff12}{$^{12}$ School of Sciences, European University Cyprus, Diogenes street, Engomi, 1516 Nicosia, Cyprus}
\noindent \affitem{aff13}{$^{13}$ {Leiden Observatory, Leiden University, P.O. Box 9513, NL-2300 RA Leiden, the Netherlands}}
\noindent \affitem{aff14}{$^{14}$ {Centro de Astrobiolog\'{i}a (CAB), CSIC-INTA, Carretera de Ajalvir km 4, Torrej\'{o}n de Ardoz 28850, Madrid, Spain}}
\noindent \affitem{aff15}{$^{15}$ {Instituto de Estudios Astrof\'{i}sicos, Facultad de Ingenier\'{i}a y Ciencias, Universidad Diego Portales, Av Ej\'{e}rcito 441, Santiago, Chile}}
\noindent \affitem{aff16}{$^{16}$ {Millenium Nucleus for Galaxies (MINGAL), Chile}}
\noindent \affitem{aff17}{$^{17}$ {Instituto de Astronom\'{i}a y Ciencias Planetarias, Universidad de Atacama, Copayapu 485, 1531772, Copiap\'{o}, Chile}}
\noindent \affitem{aff18}{$^{18}$ {INAF - Osservatorio di Astrofisica e Scienza dello Spazio di Bologna, via Gobetti 93/3, 40129, Bologna, Italy}}
\noindent \affitem{aff19}{$^{19}$ {INAF - Italian National Institute of Astrophysics, Observatory of Trieste, via G. Tiepolo 11, 34143 Trieste, Italy}}
\noindent \affitem{aff20}{$^{20}$ {Institute for Fundamental Physics of the Universe, via Beirut 2, 34151 Trieste, Italy}}
\noindent \affitem{aff21}{$^{21}$ {Kavli Institute for Cosmology, University of Cambridge, Madingley Road, Cambridge CB3 0HA, UK}}
\noindent \affitem{aff22}{$^{22}$ {Cavendish Laboratory, University of Cambridge, 19 JJ Thomson Avenue, Cambridge CB3 0HE, UK}}
\noindent \affitem{aff23}{$^{23}$ {Max-Planck-Institut f\"{u}r extraterrestrische Physik (MPE), Gie\ss{}enbachstra\ss{}e 1, 85748, Garching, Germany}}
\noindent \affitem{aff24}{$^{24}$ {Argelander-Institut f\"{u}r Astronomie, Universit\"{a}t Bonn, Auf dem H\"{u}gel 71, 53121, Bonn, Germany}}
}
}

\end{document}